\documentclass[3p,times]{elsarticle}

\usepackage{subfigure}
\usepackage{amsfonts}
\usepackage{amsthm}
\usepackage{amssymb}
\usepackage{amsmath}
\usepackage{slashed}
\usepackage{url}
\usepackage{xspace}
\usepackage{hyperref}
\usepackage{multirow}
\usepackage{color}

\usepackage{amssymb}

\usepackage[figuresright]{rotating}

\newcommand{\jpsi}{\ensuremath{J/\psi}\xspace}
\newcommand{\jpsipi}{\ensuremath{J/\psi\,\pi}\xspace}
\newcommand{\hcpi}{\ensuremath{h_c\,\pi}\xspace}
\newcommand{\etacrho}{\ensuremath{\eta_c\,\rho}\xspace}
\newcommand{\etal}{{\it et al.}}
\begin{document}

\begin{frontmatter}




\title{Probing the nature of $Z_c^{(\prime)}$ states via the $\etacrho$ decay}


\author[pippo]{A. Esposito}
\author[pappo]{A.L. Guerrieri}
\author[sempronio]{A. Pilloni}
\address[pippo]{Department of Physics, 538W 120th Street, Columbia University, New York, NY, 10027, USA}
\address[pappo]{Dipartimento di Fisica and INFN, Universit\`a di Roma ``Tor Vergata'', Via della Ricerca Scientifica 1, I-00133 Roma, Italy}
\address[sempronio]{Dipartimento di Fisica and INFN, ``Sapienza'' Universit\`a di Roma, P.le A. Moro 2, I-00185 Roma, Italy}

\begin{abstract}
The nature of the so-called $XYZ$ states is a long-standing problem. It has been suggested that such particles may be described as compact four-quark states or loosely bound meson molecules. In the present work we analyze the $Z_c^{(\prime)}\to\etacrho$ decay using both approaches. Such channel might provide useful insights on the nature of the $Z_c^{(\prime)}$, helping discriminating between the two different models.
\end{abstract}

\begin{keyword}
Exotic charmonium \sep Tetraquarks \sep Meson Molecules
%
\PACS 14.40.Rt \sep 12.39.Jh \sep 12.39.Hg
\end{keyword}

\end{frontmatter}


\section{Introduction}
\label{intro}Since the first observation of the $X(3872)$, made by Belle in 2003~\cite{xbelle}, a considerable number of ``exotic'' particles has been discovered in the heavy quark sector (for a review see~\cite{review}). In particular, the  finding of charged charmonium-like resonances is a compelling evidence of the fact that these resonances cannot be fitted into the known frameworks.
However, their nature still lacks of a comprehensive theoretical understanding. The most accepted phenomenological models interpret them as compact tetraquark states~\cite{maianiX}, as loosely bound meson molecules~\cite{mols,cleven}, as a quarkonium state interacting with light matter via residual strong forces~\cite{hadrocharmonium}, or as hybrid gluonic states~\cite{hybrids}.

In 2013, the BES and Belle collaborations announced the discovery of a charged charmonium-like state called $Z_c(3900)$~\cite{zc}, in the channel $Y(4260)\to Z_c(3900)^+ \pi^- \to \jpsi \pi^+ \pi^-$. Soon after, evidence for the channel $Z_c(3900)^+\to (DD^*)^+$ was reported by BES~\cite{zcDDstar}. In the meanwhile, BES found a twin $Z_c^\prime(4020)$ state decaying into $h_c\, \pi^+$ and $\bar D^{*0} D^{*+}$~\cite{zc4020}. The most likely quantum numbers for both resonances are $(I^G)J^{PC} = (1^+) 1^{+-}$~\footnote{For $C$ we report the eigenvalue of the neutral isospin partner.}.

These two states happen to be close to the $DD^*$ and $D^*D^*$ thresholds, respectively, which might suggest a molecular interpretation~\cite{riddle}. However, the mechanism for which these would-be-molecules are pushed slightly above the threshold (by tens of MeV) is still unclear. On the other hand, the constituent diquark-antidiquark model predicted the existence of a $1^{+-}$ resonance around $3882$~MeV~\cite{maianiX,maianizc}, and of its radial excitation around $4470$~MeV. The confirmation by the LHCb collaboration of a charged $Z(4430)$ state decaying into $\psi(2S)\,\pi^+$~\cite{z4430lhcb} and far from open charm thresholds strengthens the whole picture~\cite{pioni}.
The $Z_c^\prime(4020)$ lacked an interpretation in the original model, but this is fixed by a recent ``type~II'' tetraquark paradigm~\cite{maianitypeII}. For a discussion on the production of these states at hadron colliders, see~\cite{montecarlo,pioni}. 

The Belle collaboration recently started an analysis with the aim of searching for such exotic resonances decaying into the $\eta_c$ charmonium~\cite{Belleweb}. A possible interesting channel could be $Z_c^{(\prime)}\to \etacrho$, as it can provide a good way to discriminate between the interpretations mentioned before. In Section~\ref{tetra} we will discuss this decay channel according to the main tetraquark models. In Section~\ref{molecular} we will evaluate the branching fractions according to the molecular hypothesis, by means of a non-relativistic effective theory. Our conclusions are in Section~\ref{conclusions}.

\section{The compact tetraquark}
\label{tetra}In the constituent diquark-antidiquark models, the ground state tetraquarks
are the eigenstates of the color-spin Hamiltonian:
\begin{equation}
  H= \sum_i m_i -2 \sum_{i \neq j,a} \kappa_{ij} \vec S_i \cdot \vec S_j \frac{\lambda^a_i}{2} \cdot \frac{\lambda^a_j}{2}.\label{eq:tetraham}
\end{equation}
\begin{figure}[t]
\centering
\includegraphics[width=.5\textwidth]{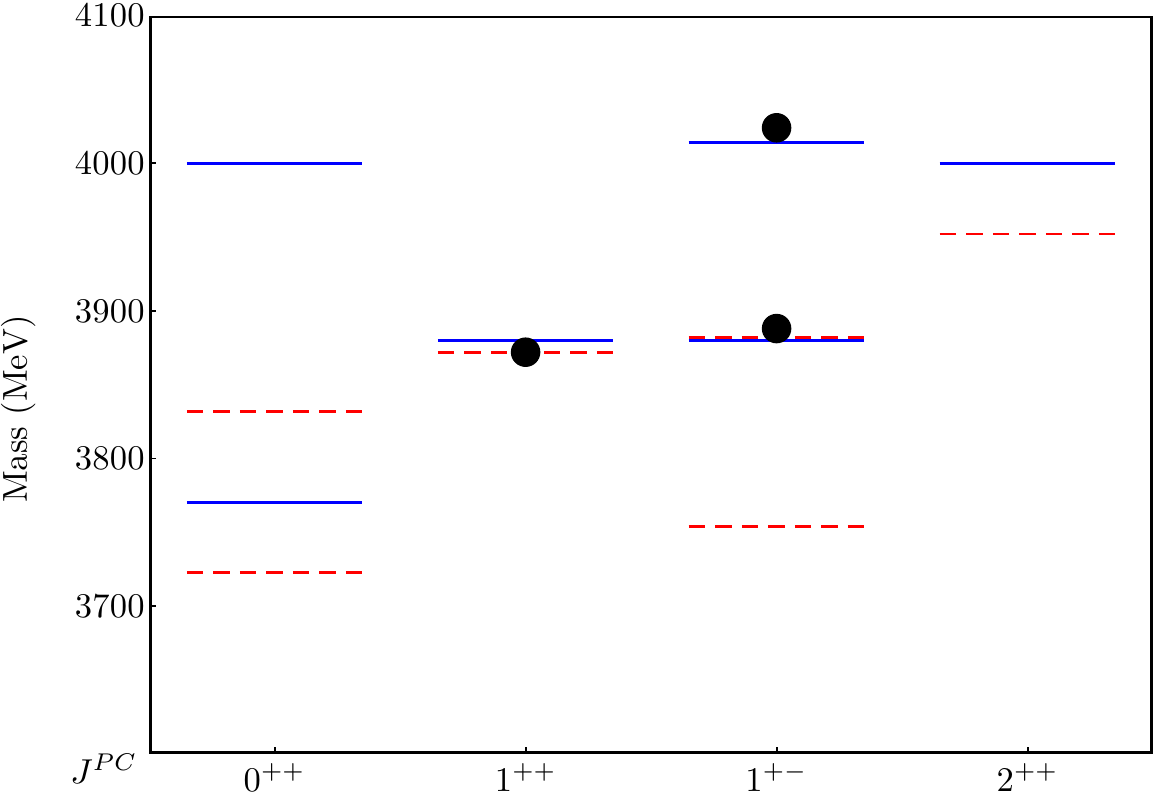}
\caption{Levels predicted by the type~I (red dashed)~\cite{maianiX} and type~II (solid blue)~\cite{maianitypeII} tetraquark models. The black disks are the experimental $X(3872)$, $Z_c(3900)$ and $Z_c^\prime(4020)$ masses.}
\label{fig:duemodelli}
\end{figure}
\begin{table}[b]
\centering
 \begin{tabular}{l|c|c|c|c|c|c}\hline\hline
 & \multicolumn{3}{c|}{Diquark basis} & \multicolumn{3}{c}{Closed charm basis} \\ \hline
   & $\left|1_{cq},0_{\bar c \bar q}\right\rangle$ & $\left|0_{cq},1_{\bar c \bar q}\right\rangle$ & $\left|1_{cq},1_{\bar c \bar q}\right\rangle$ & 
  $\left|1_{c\bar c},0_{q \bar q}\right\rangle$  & $\left|0_{c\bar c},1_{q \bar q}\right\rangle$ & $\left|1_{c \bar c},1_{q \bar q}\right\rangle$ \\ \hline\hline
 $X\left( 1^{++} \right)$ & $1/\!\sqrt{2}$  & $1/\!\sqrt{2}$ & 0 & 0 &  0 &  1 \\ \hline
 $Z\left( 1^{+-} \right)$ & $\cos \theta/\!\sqrt{2}$ & $-\cos \theta/\!\sqrt{2}$ & $\sin\theta$ & $\sin\left( \theta+45^\circ\right)$ & $-\cos\left( \theta+45^\circ\right)$ & 0 \\ \hline
 $Z^\prime \left( 1^{+-} \right)$ & $-\sin \theta/\!\sqrt{2}$ & $\sin \theta/\!\sqrt{2}$ & $\cos\theta$ & $\cos\left( \theta+45^\circ\right)$ & $\sin\left( \theta+45^\circ\right)$ & 0 \\ \hline
 \end{tabular}
 \caption{Quark content of the axial tetraquarks respectively in the diquark and closed charm basis. The symbol $S_{q_1 q_2}$ stands for the spin eigenstate of the $q_1 q_2$ pair. \mbox{In type~I model, $\theta \simeq -47^\circ$. In type~II model, $\theta = 0^\circ$.}}
\label{tab:states}
 \end{table}

In the type~I model, the $\kappa_{ij}$ coefficents are extracted from ordinary meson
and baryon masses, and the $X(3872)$ mass is used as input to fit the diquark
mass~\cite{maianiX}. The predicted mass spectrum is reported in \figurename{~\ref{fig:duemodelli}} (dashed lines). The mixing angle of the two $1^{+-}$ states is found to be $\theta \simeq -47^\circ$. The spin-$1$ eigenstates $X$ and $Z$ in \tablename{~\ref{tab:states}} are identified as the $X(3872)$ and the $Z_c(3900)$, predicting a mass of $3882$ MeV for the latter.  The $Z^\prime\left( 1^{+-}\right)$ state would have a mass of $3755$ MeV, but no resonances have been found with this mass and quantum numbers. 

Recently, it was proposed to neglect the spin-spin interaction outside
the diquarks (type~II)~\cite{maianitypeII}. More specifically, in~\eqref{eq:tetraham} all couplings but $\kappa_{cq} = \kappa_{\bar c \bar q}$ are neglected. The resulting spectrum contains two degenerate levels with quantum numbers $1^{++}$ and $1^{+-}$, which can be identified as the $X$ and the $Z_c(3900)$, and a heavier $1^{+-}$ state corresponding to the $Z_c^\prime(4020)$. This hamiltonian is diagonal in the diquark-antidiquark base, hence the wave functions in \tablename{~\ref{tab:states}} have $\theta = 0^\circ$. The mass spectrum of the latter model is reported in \figurename{~\ref{fig:duemodelli}} (solid lines).

As far as the transition matrix elements are concerned, we can consider that the heavy spinor structure $\chi_c$ factorizes out from the total wave function $\psi_c=\chi_c\otimes \phi_c\!\left(\vec r\right)$ in the infinite quark mass limit, thus realizing heavy quark spin symmetry~\cite{hqss,guohqss}. The correction to this are suppressed by inverse powers of the heavy mass itself. 
This applies to any state containing a heavy quark, and hence to our diquarks as well. In particular, the factorization of the diquark wave function $\psi_{[cq]}=\chi_c\otimes \phi_{[cq]}\!\left(\vec r_c, \vec r_q, s_q\right)$ is expected to have corrections of order $\Lambda_{QCD}/m_c\sim 0.25$~\cite{hqss}.
A similar expansion also holds for the transition operator allowing the decay into charmonia, which at leading order can be expressed as a direct sum, 
$\mathbf{1}_{HS} \oplus \hat{T}_{\perp HS}$, the first operator preserving the spin of the heavy pair.
At leading order, we can consistently use the unperturbed wave functions to evaluate the decay amplitudes, obtaining
\begin{equation}
\mathcal{A}=\langle \chi_{c\bar c}|\chi_c \otimes \chi_{\bar c}\rangle \langle \phi_{c\bar c} | \hat{T}_{\perp HS} | \phi_{[cq][\bar c \bar q]} \rangle+\mathcal{O}\left(\frac{\Lambda_{QCD}}{m_c}\right).
\label{eq:succzz}
\end{equation}
The first term is simply a Clebsch-Gordan coefficient enforcing heavy quark spin symmetry, and can be evaluated upon a Fierz transformation.
The second term depends on the actual dynamic of the decay. For each tetraquark state, one usually assumes this term to be the same for every charmonium, and therefore cancels out when evaluating ratios of decay widths~\cite{maianitypeII,hqss,guohqss}.
In other words, it is possible to
re-arrange the $c \bar c$ quarks into a color singlet pair with a Fierz
transformation (see \tablename{~\ref{tab:states}}), and consequently the effective coupling of the tetraquark
to a charmonium can be assumed to be proportional to the $c \bar c$ component with the
appropriate spin content, with an expected error $\mathcal{O}(\Lambda_{QCD}/m_c)$. 

\begin{figure}[t]
\centering
\includegraphics[width=.5\textwidth]{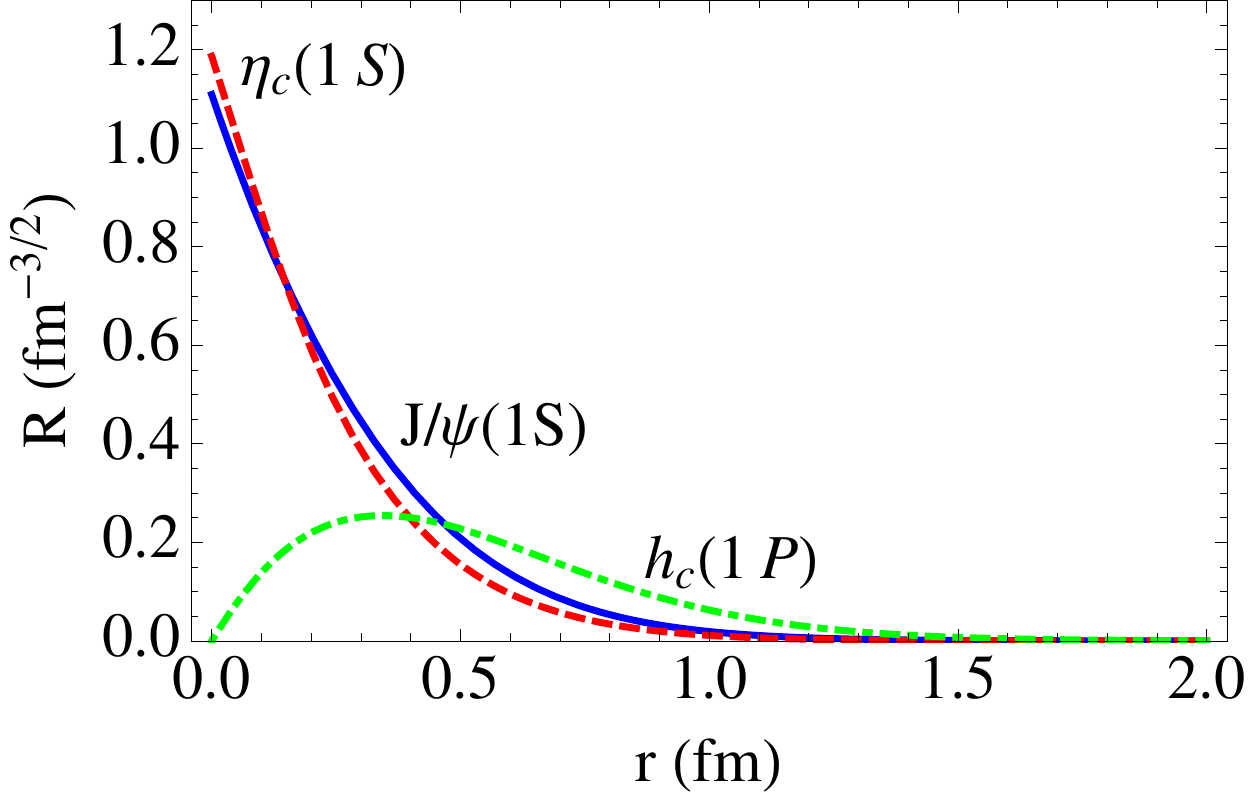}
\caption{Radial component of the solutions of the Schr\"odinger equation in presence of the full Cornell potential~\eqref{eq:cornell}. The dashed red line is the radial wave function of the $\eta_c$, the solid blue line 
is the wave function of the $\jpsi$ and the dash-dotted green line is the wave function of the $h_c$.}
\label{fig:charm}
\end{figure}

The kinematics of the transition matrix element in Eq.~\eqref{eq:succzz} can be taken into account parametrizing the matrix elements as an effective coupling times the most general Lorentz-invariant combination of polarization vectors and momenta with appropriate behaviour under parity and charge conjugation~\cite{maianizc}, \emph{i.e.}:
\begin{subequations}
\begin{gather}
 \left\langle \jpsi\left(\eta,p\right)\,\pi\left(q\right) | Z\left(\lambda,P\right) \right\rangle = g_{Z\psi\pi}\, \lambda \cdot \eta,\quad \left\langle \eta_c\left(p\right)\,\rho\left(\epsilon,q\right) | Z\left(\lambda,P\right) \right\rangle = g_{Z\eta_c\rho} \,\lambda \cdot \epsilon,\\  \left\langle h_c\left(p,\eta\right)\,\pi\left(q\right) | Z\left(\lambda,P\right) \right\rangle = \frac{g_{Z h_c\pi}}{M_Z^2}\, \epsilon^{\mu\nu\rho\sigma}\lambda_\mu \eta_\nu P_\rho q_\sigma,
\end{gather}
\end{subequations}
where $\eta$, $\lambda$ and $\epsilon$ are the polarization vectors, $p$, $q$ and $P$ are the momenta and the $g$s are the effective couplings  with the dimension of a mass. Hence, the partial decay widths are:
\begin{subequations}
\begin{gather}
\Gamma\left(Z\to\jpsi \pi\right) = \frac{1}{3} \frac{p^*}{8\pi M_Z^2} g_{Z\psi\pi}^2 \left(3 + \frac{p^{*2}}{M_{\jpsi}^2}\right),\quad
\Gamma\left(Z\to\eta_c \rho\right) = \frac{1}{3} \frac{p^*}{8\pi M_Z^2} g_{Z\eta_c\rho}^2 \left(3 + \frac{p^{*2}}{M_\rho^2}\right),\\
\Gamma\left(Z\to h_c \pi\right) = \frac{1}{3} \frac{p^*}{8\pi M_Z^2} g_{Z h_c \pi}^2 \left( 2\frac{p^{*2}}{M_Z^2} \right).
\end{gather}
\end{subequations}
where $p^*$ is the decay 3-momentum in the $Z$ rest frame. 
The original models neglect the spatial dependence of the wave functions:
the coupling to each heavy quark spin configuration is somewhat
universal, and the differences are only of kinematical nature. Hence, the $P$-wave decays into $h_c\,\pi$ are highly suppressed
by phase space. Moreover, in the type~I model the $0_{c\bar c}$ component of $Z_c(3900)$ is $\cot 2^\circ \simeq30$ times larger than the $1_{c\bar c}$ one, thus predicting a peak in \etacrho much more intense than the one in the discovery channel \jpsipi.

A different assumption on the $\langle \phi_{c\bar c} | \hat{T}_{\perp HS} | \phi_{[cq][\bar c \bar q]} \rangle$ matrix element has been recently proposed by Brodsky \etal~\cite{Brodsky:2014xia}, in order to take into account the spatial wave functions of the involved states. A diquark-antidiquark pair would tend to convert all its kinetic energy into potential energy of the color flux tube until it can be considered at rest at a fixed relative distance which satisfies $V(r_Z, S_1\!=\!S_2\!=\!0) = M - 2m_{cq}$, where $m_{cq} = (1.86\pm0.10)\text{~GeV}$~\cite{Brodsky:2014xia} is the constituent diquark mass (evaluated via QCD sum rules), and $V(r)$ is the Cornell potential:

\begin{equation}
V(r)=-\frac{4}{3}\frac{\alpha_s}{r} + br + \frac{32\pi \alpha_s}{9 m_c^2} e^{-\sigma^2 r^2} \vec S_1 \cdot \vec S_2,\label{eq:cornell}
\end{equation}
where $\alpha_s=0.5461$, $b=0.1425$ GeV$^2$, $m_c=1.4797$~GeV and $\sigma = 1.0946$~GeV$^2$~\cite{Brodsky:2014xia} are the parameter which better describe the charmonium spectrum.
Thus, the larger a charmonium wave function is at $r_Z$, the more favored the decay into such state will be~\footnote{Light mesons are not expected to play a role in this mechanism, having a much broader wave function. The argument is someway similar to the hadrocharmonium theory~\cite{hadrocharmonium}.}. In other words, we can assume the effective coupling to be proportional to $\left|\psi_{c \bar c}\left(r_Z\right)\right|^2$.
The charmonium wave functions can be evaluated by solving the Schr\"odinger equation with a Cornell potential and the adequate spin assignment.
Concretely, we find $r_{Z_c}=(0.60\pm0.17)$~fm and $r_{Z_c^\prime}=(0.72\pm0.19)$~fm
and estimate from the ratios of charmonium radial wave functions (\figurename{~\ref{fig:charm}}):
\begin{equation}
\frac{g^2_{Z \eta_c \rho}}{g^2_{Z \psi\pi}}=0.68_{-0.12}^{+0.15}; \quad \frac{g^2_{Z^\prime \eta_c\rho}}{g^2_{Z^\prime h_c\pi}}=\left(5.7_{-4.5}^{+24.4}\right)\times 10^{-2}.
\label{eq:ratiocoupling}
\end{equation}

In \tablename{~\ref{tab:tetra}} we report the predicted branching fractions to \etacrho, according to the presented models. Here and in the rest of the paper, the uncertainties are estimated using a toy Monte Carlo simulation. In particular, the multiplicative relative errors are extracted from a log-normal distribution to ensure positivity, while all the other quantites are otherwise assumed to be Gaussian. The asymmetric errors will always report the $68\%$ C.L. with respect to the mode of the likelihood distribution. Despite the proposed mechanism for a dynamical description is still preliminary, the values obtained are not significantly different from a pure kinematical estimates, which give more strength to our predictions.

 \begin{table}[h]
\centering
 \begin{tabular}{l|c|c|c|c}\hline\hline
 & \multicolumn{2}{c|}{Kinematics only} & \multicolumn{2}{c}{Dynamics included} \\ \hline
  & type~I & type~II & type~I & type~II \\ \hline\hline
\rule{0pt}{20pt} $\displaystyle{\frac{\mathcal{BR}\left(Z_c \to \etacrho\right)}{\mathcal{BR}\left(Z_c \to \jpsipi\right)}}$ & $\left(3.3^{+7.9}_{-1.4}\right)\times 10^2$ & $0.41^{+0.96}_{-0.17}$ & $\left(2.3^{+3.3}_{-1.4}\right)\times10^2$ & $0.27^{+0.40}_{-0.17}$ \\[10pt] \hline
\rule{0pt}{20pt} $\displaystyle{\frac{\mathcal{BR}\left(Z_c^\prime \to \etacrho\right)}{\mathcal{BR}\left(Z_c^\prime \to h_c \pi\right)}}$ & \multicolumn{2}{c|}{$\left(1.2^{+2.8}_{-0.5}\right)\times10^2$} & \multicolumn{2}{c}{$6.6^{+56.8}_{-5.8}$} \\[10pt] \hline 
 \end{tabular}
 \caption{Predicted ratios of branching ratios for $Z_c^{(\prime)}$ states according to the main tetraquark models. The results in the first column are computed considering kinematics and heavy quark spin symmetry only. In the second column, instead, we estimated them using the method of~\cite{Brodsky:2014xia}. Both type~I and type~II model give the same predictions for the $\mathcal{BR}\left(Z_c^\prime \to \etacrho\right)/\mathcal{BR}\left(Z_c^\prime \to h_c \pi\right)$, since both $h_c$ and $\eta_c$ have $s_{c\bar c}=0$. The ratio of the branching fractions into  $\etacrho$ and $\jpsipi$ according to the type~I model are so large because of the prefactor $\cot^2 2^\circ \simeq 900$. The errors are estimated via the toy MC simulation as explained in the text.}
\label{tab:tetra}
 \end{table}

\section{The molecular picture}
\label{molecular}In the meson molecule model, the $Z^{(\prime)}_c$ is interpreted as a $\left(D^*\bar{D}^{(*)}\right)_{C=-1}$ resonant state. We will evaluate the branching fraction $Z^{(\prime)}_c \to \etacrho$ by means of the Non-Relativistic Effective Field Theory (NREFT)~\cite{cleven}, a framework based on HQET and NRQCD. The interaction vertices between the molecular $Z_c^{(\prime)}$, the $\eta_c$ and the $D^{(*)}$ mesons have been discussed in~\cite{cleven}, and are given by:
\begin{equation}
\mathcal{L}_{Z_c^{(\prime)}}=\frac{z^{(\prime)}}{2}\left\langle \mathcal{Z}^{(\prime)}_{\mu,ab} \bar H_{2b}\gamma^\mu\bar H_{1a}\right\rangle+h.c.,
\end{equation}
where $\langle\,\cdots\rangle$ stands for a trace over Dirac indices and $a$ and $b$ are isospin indices. The details on the definitions of the HQET superfields and our conventions are reported in~\ref{appA}. 
This kind of HQET Lagrangian could describe the decays of $Z_c$'s regardless of the internal structure of such states, and provide model-independent results in terms of unknown effective couplings. In these effective theories the molecular nature  is taken into account by forcing these states to couple to their own constituents only, and forbidding all other tree level vertices. The transition to charmonia would thus be allowed only by heavy meson loops.

The interaction between heavy mesons and charmonia is given by~\cite{Colangelo:2003sa,cleven}:
\begin{equation} \label{Lpsi}
\mathcal{L}_{c\bar c} =\frac{g_2}{2}\left\langle\bar \Psi H_{1a} \overleftrightarrow{\slashed{\partial}} H_{2a}\right\rangle 
+\frac{g_1}{2} \left\langle \bar \chi_\mu H_{1a} \gamma^\mu H_{2a}\right\rangle +h.c.,\\
\end{equation}
where $\Psi$ is the superfield containing the $S$-wave charmonia, and $\chi$ the superfield of the $P$-wave charmonia. 
The couplings $g_1$ and $g_2$ have been estimated to be $g_1=-\sqrt{M_{\chi_{c0}}/3}\big/f_{\chi_{c0}}$ and $g_2=\sqrt{M_{\jpsi}}\big/2M_Df_{\jpsi}$~\cite{Colangelo:2003sa}, where $f_{\chi_{c0}}=(510\pm40)$~MeV is estimated via QCD sum rules, and $f_{\jpsi}=(405\pm14)$~MeV can be obtained from the electronic width of the $\jpsi$ itself~\footnote{The same formula can be used also to evaluate the coupling $g_2^\prime$ of the $\psi(2S)$, giving $g_2^\prime \sim 1.5 \,g_2$.}. Plugging all the values one gets $g_1=(-2.09\pm0.16)$~GeV$^{-1/2}$ and $g_2=(1.16\pm0.04)$~GeV$^{-3/2}$.

\begin{figure}[t]
\centering
\subfigure[]{ \label{fig:1a}
\includegraphics{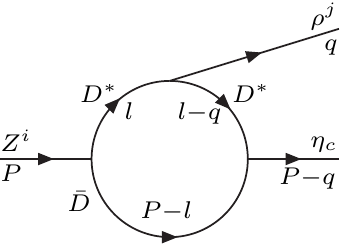}
}
\hspace{.5em}
\subfigure[]{ \label{fig:1b}
\includegraphics{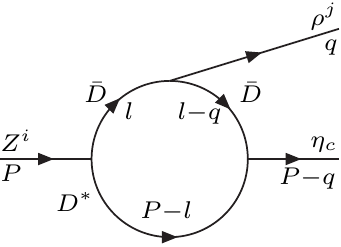}
}
\hspace{.5em}
\subfigure[]{ 
\label{fig:1c}
\includegraphics{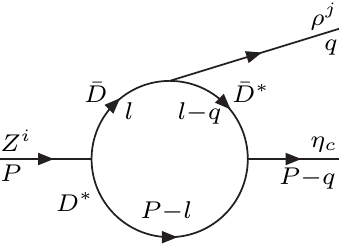}
}
\caption{Possible one-loop diagrams for $Z_c\to\etacrho$. The charge conjugate diagrams are omitted.} \label{fig:one_loop}
\end{figure}

One key ingredient of our analysis is the interaction between heavy mesons and light mesons. The relevant HQET Lagrangian can be found in~\cite{Casalbuoni:1996pg}. We report here the terms involving two heavy mesons and one or two $\rho$:
\begin{align}\label{eq:Lrho}
\mathcal{L}_{\rho DD^*}=&i\beta\left\langle H_{1b} v^\mu\left(\mathcal{V}_\mu-\rho_\mu\right)_{ba} \bar H_{1a}\right\rangle+i\lambda\left\langle H_{1b}\sigma^{\mu\nu} F_{\mu\nu}(\rho)_{ba} \bar H_{1a}\right\rangle+h.c.,
\end{align}
where $F_{\mu\nu}(\rho)=\partial_\mu\rho_\nu-\partial_\nu\rho_\mu+[\rho_\mu,\rho_\nu]$ and $\sigma^{\mu\nu}=i[\gamma_\mu,\gamma_\nu]/2$. The vector field $\mathcal{V}_\mu$ contains information on pion pairs, which are not of interest for the present case.
The imposition of Vector Meson Dominance implies $\beta=0.9 \pm 0.1$. QCD sum rules give $\lambda=(0.56\pm0.07)$ GeV$^{-1}$~\cite{Isola:2003fh}. The errors on these two couplings have been estimated in~\cite{Li:2007dv}, where authors found compatible values using QCD sum rules.

\subsection{Power counting}
The relevant one-loop diagrams are reported in \figurename{~\ref{fig:one_loop}}. As shown in~\cite{cleven}, the importance of a certain diagram in NREFT can be estimated using a power counting procedure. The heavy meson velocity relevant for the decay/production of some particle $X$ can be evaluated as $v_X\!\sim\!\sqrt{\left|M_X-2M_D\right|/M_D}$, which in our case gives $v_Z\simeq 0.12$ and $v_\eta\simeq 0.68$. Every meson loop counts as $v_X^5/{(4\pi)}^2$, while the heavy meson propagator scales as $1/v_X^2$. In case there is more than one heavy quark external line joined to a loop, we will use an average velocity $\bar v$. Moreover, depending on the possible presence of derivatives in the interaction vertex, the diagram may also scale either as a power of the external momentum of the $\rho$ meson, $q\simeq426.7$ MeV, or as an additional power of $v_X$. According to these rules, the combination of the diagrams in \figurename{~\ref{fig:one_loop}} scale as:
\begin{equation}
\frac{\bar v^5}{{(4\pi)}^2}\frac{1}{\bar v^6} \beta g_V \frac{q}{m_{\rho}} \frac{q}{M_D}\simeq 1\times10^{-2} \quad\text{or}\quad \frac{\bar v^5}{{(4\pi)}^2}\frac{1}{\bar v^6} \lambda g_V q\frac{q}{M_D}\simeq 5\times10^{-3}. \end{equation}
Powers of $M_D$ have been introduced to make everything dimensionless.

\begin{figure}[b]
\centering
\subfigure[]{ \label{fig:2loopa}
\includegraphics{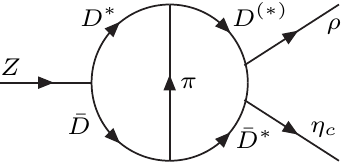}
}
\subfigure[]{ \label{fig:2loopb}
\includegraphics{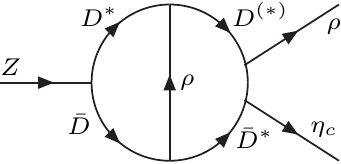}
}
\subfigure[]{ \label{fig:2loopc}
\includegraphics{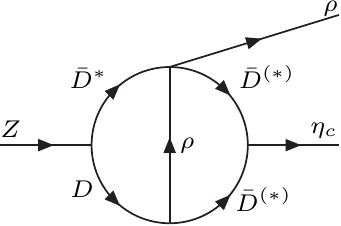}
}
\caption{Most relevant two-loop contributions.} \label{fig:2loop}
\end{figure}

We now need to evaluate the contribution from higher number of loops since it may happen for them to be as relevant as the one-loop contributions~\cite{cleven}. The possible processes we can have are reported in Fig.~\ref{fig:2loop}. In particular, the internal virtual particle can be either a pion or another $\rho$ meson. Each pion vertex counts as $gp_\pi/F_\pi$, where $g\simeq0.5$ is the axial coupling, $p_\pi$ is the pion momentum and $F_\pi\simeq93$~MeV 
is the pion decay constant. Since $m_\pi\simeq0$ the pion propagator scales a $1/M_D^2v_\eta^2$, $M_Dv_\eta$ being the largest momentum running inside the loop -- see~\cite{cleven}. The other two diagrams contain, instead, an additional $\rho$ meson running in the internal leg. In particular, in \figurename{~\ref{fig:2loopc}} we employed the $\rho\rho D^{(*)}D^{(*)}$ vertex arising from the chiral Lagrangian in Eq.~\eqref{eq:Lrho}. This new vertex does not contain any derivative and hence counts as $\lambda g_V^2$. Moreover, since the mass of the $\rho$ cannot be neglected, the internal propagator scales as $1/(M_D^2v_\eta^2+m_\rho^2)$. These three diagrams count respectively as:
\begin{subequations}
\begin{gather}
\frac{v_Z^5}{{(4\pi)}^2}\frac{1}{v_Z^4}\frac{v_\eta^5}{{(4\pi)}^2}\frac{1}{v_\eta^4}\frac{1}{M_D^2v_\eta^2}\frac{g^2 \,\beta\, g_V}{F_\pi^2 \,m_\rho}q^4 M_D\simeq3\times10^{-5},  \quad  \frac{v_Z^5}{{(4\pi)}^2}\frac{1}{v_Z^4}\frac{v_\eta^5}{{(4\pi)}^2}\frac{1}{v_\eta^4}\frac{1}{M_D^2v_\eta^2+m_\rho^2}\frac{\lambda\,\beta\,g_V^3}{m_\rho} q^4 M_D\simeq 7\times 10^{-6} \\
\text{and }\quad \frac{v_Z^5}{{(4\pi)}^2}\frac{1}{v_Z^4}\frac{v_\eta^5}{{(4\pi)}^2}\frac{1}{v_\eta^4}\frac{M_D^2}{M_D^2v_\eta^2+m_\rho^2}\lambda \beta g_v^3\frac{q^2}{m_\rho}\simeq2\times10^{-4}. 
\end{gather}
\end{subequations}
As one can see, the first two kinds of diagrams are at least two orders of magnitude smaller than the one-loop diagrams and hence can be neglected. The dominant two-loop contribution is thus given by the diagrams like the one in \figurename{~\ref{fig:2loopc}}. There are 6 diagrams of this kind (depending on which $D$ mesons run in the loop), while there are 3 one-loop diagrams. Therefore, the ratio between the two-loop and the one-loop contributions is $(6\times2\times10^{-4})/(3\times5\times10^{-3})\simeq0.08$. To be conservative, we will assign a $15\%$ relative error to each single amplitude.

\begin{figure}[t]
\centering
\subfigure[]{ \label{fig:1a_prime}
\includegraphics{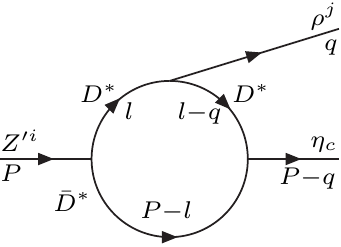}
}
\hspace{.5em}
\subfigure[]{ \label{fig:1b_prime}
\includegraphics{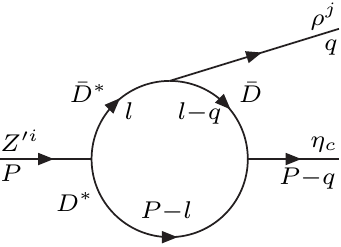}
}
\caption{Possible one-loop diagrams for $Z_c^\prime \to\etacrho$. The charge conjugate diagrams are omitted. 
} \label{fig:one_loop_prime}
\end{figure}

The case of $Z_c^\prime$ is very close to the one presented here and we will thus omit it. However, it is worth noting that, since its constituents are $D^*\bar D^*$, there is one diagram less at the lowest order (\figurename{~\ref{fig:one_loop_prime}}).


\subsection{Branching fractions}
We now have all the tools to evaluate the ratio between the branching fractions $\mathcal{BR}(Z_c\to\etacrho)/\mathcal{BR}(Z_c\to \jpsipi)$ and $\mathcal{BR}(Z_c^\prime\to\etacrho)/\mathcal{BR}(Z_c^\prime\to \hcpi)$. The amplitudes for the $Z_c^{(\prime)}\to J/\psi(h_c)\,\pi$ processes are taken from~\cite{cleven} while the ones for $Z_c^{(\prime)}\to\eta_c\rho$ are reported in \ref{appB}. We obtain:
\begin{align}
\frac{\mathcal{BR}(Z_c\to\etacrho)}{\mathcal{BR}(Z_c\to \jpsipi)} = \left(4.6_{-1.7}^{+2.5}\right)\times 10^{-2}\,;\quad\frac{\mathcal{BR}(Z_c^\prime\to\etacrho)}{\mathcal{BR}(Z_c^\prime\to\hcpi)}=\left(1.0_{-0.4}^{+0.6}\right)\times 10^{-2}\,.
\end{align}
Moreover, the widths have been computed with: 
\begin{align}
\Gamma(Z\to B+C)=\frac{1}{3} \frac{p^*}{8\pi\, m_Z^2} \sum_\text{pol}\left|\mathcal{A}\right|^2.
\end{align}
Both $Z_c\to\eta_c\rho$ and $Z_c\to J/\psi\pi$ processes contain two $P$-wave vertices, however the first one has a smaller phase space that, by itself, leads to an order of magnitude suppression. The dominance of $Z_c^\prime\to h_c\pi$ over $Z_c^\prime\to\eta_c\rho$, instead, is explained by the presence of the extra $\eta_cDD^*$ $P$-wave vertex in the latter process. 

Other interesting ratios can be evaluated. 
We can assume the total width to be saturated by the decays into $D^{(*)}D^*$, $\etacrho$, $\hcpi$, $\jpsipi$, $\psi(2S)\pi$, and use the measured widths~\cite{pdg} to extract the coupling of the molecular $Z_c^{(\prime)}$ to its constituents:
\begin{align}
\left|\,z\,\right|=\left(1.26_{-0.14}^{+0.14}\right) \text{ GeV}^{-1/2}\quad\text{ and }\quad\left|\,z^\prime\,\right|= \left(0.58_{-0.19}^{+0.22}\right) \text{ GeV}^{-1/2},
\end{align}
where the uncertainties are due to both the experimental uncertainties on the total widths and to the theoretical uncertainties arising from neglecting the two-loop diagrams. In particular, the relative errors for the different channels has been estimated starting from the power counting performed in~\cite{cleven} and replacing the right quantities for the charm sector. In doing so we found the following conservative uncertainties: $15\%$ on $h_c\pi$, $25\%$ on $\jpsipi$ and $10\%$ on $\psi(2S)\pi$. The $DD^*$ channel has no theoretical error, being a tree-level process. Note the relevant amount of spin symmetry violation, much larger than in the bottom sector~\cite{cleven}. Using again the amplitudes computed in \cite{cleven} -- the relevant Lagrangian being the one reported in Eq.~\eqref{Lpsi} -- and the PDG values for the total widths~\cite{pdg}, we can obtain the ratios we are looking for:
\begin{align}
\frac{\mathcal{BR}(Z_c\to \hcpi)}{\mathcal{BR}(Z_c^\prime\to \hcpi)}=0.34_{-0.13}^{+0.21}\,;\quad\frac{\mathcal{BR}(Z_c\to \jpsipi)}{\mathcal{BR}(Z_c^\prime\to \jpsipi)}=0.35_{-0.21}^{+0.49}\,.
\end{align}

While this work was written up, a paper discussing the decays $Z_c^{(\prime)}\to \etacrho$ in an Effective Lagrangian approach appeared~\cite{cinese}, quite in agreement with our results.

\section{Conclusions}\label{conclusions}
In this letter we discuss the possible decays $Z_c^{(\prime)}\to\etacrho$ within the most accepted phenomenological models. 

\begin{figure}[t]
\centering
\includegraphics[width=.49\textwidth]{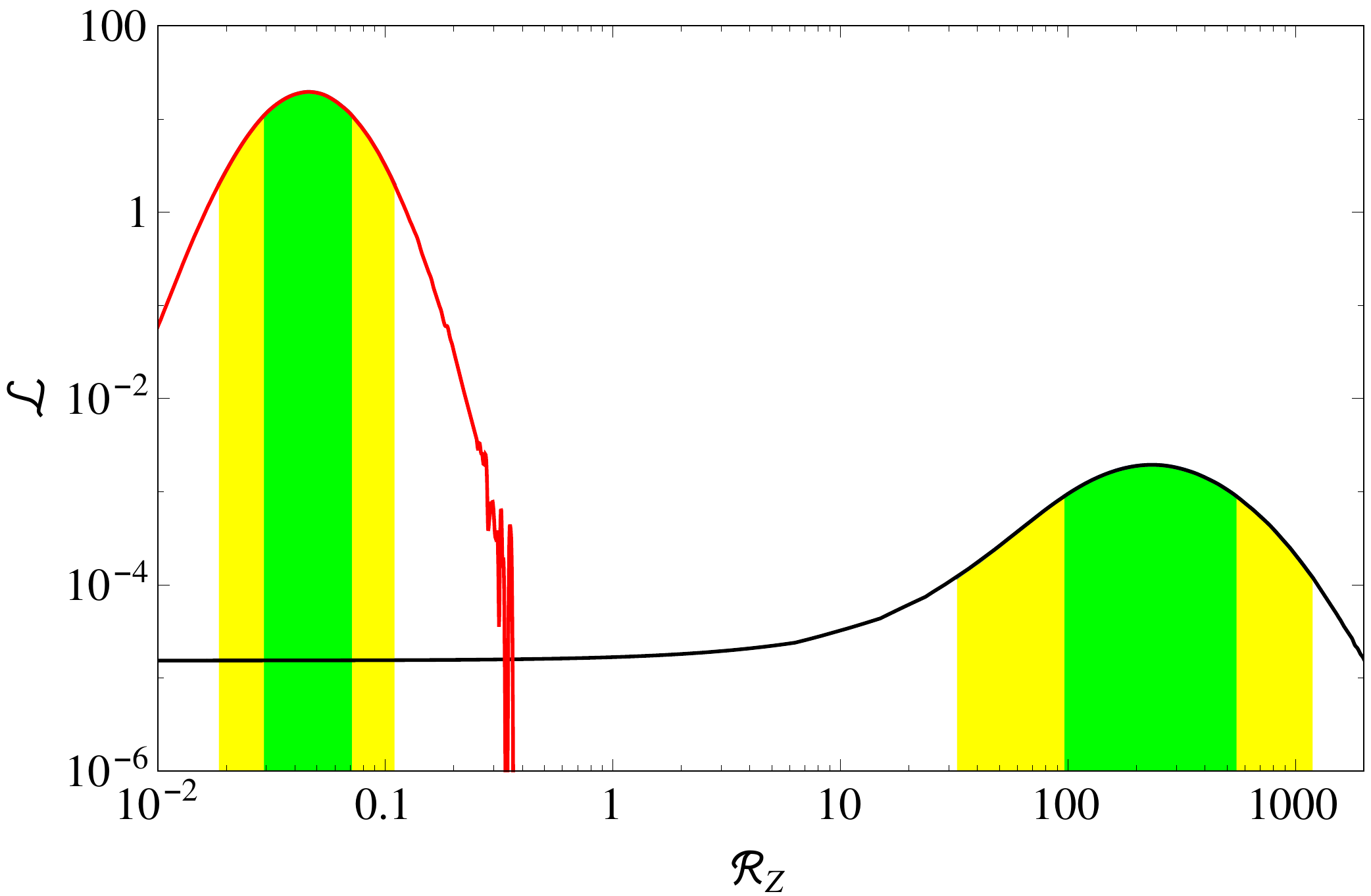}
\includegraphics[width=.49\textwidth]{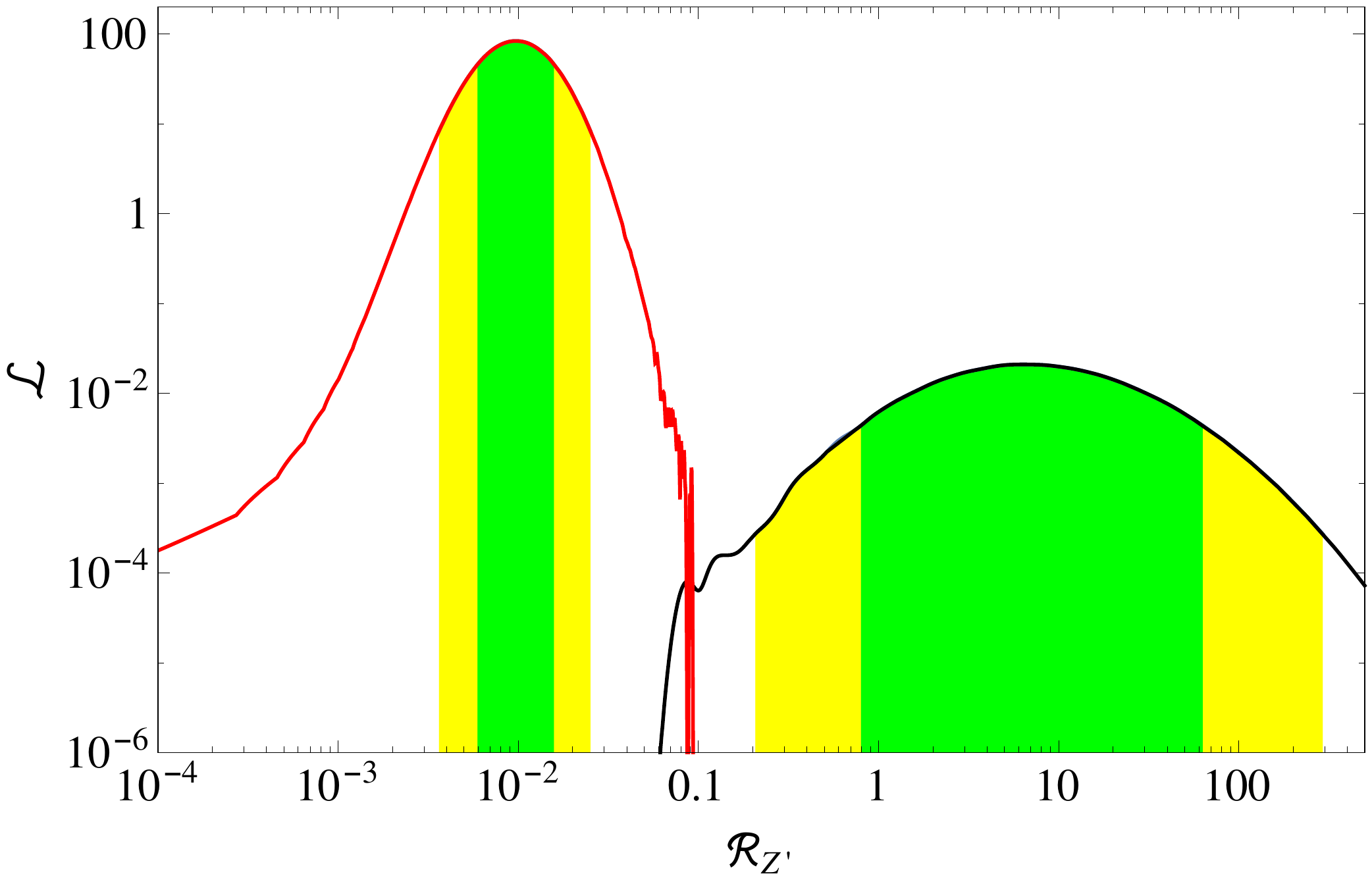}
\caption{Likelihood curves for $\mathcal{BR}(Z_c\to \etacrho)/\mathcal{BR}(Z_c\to \jpsipi)$ (left) and $\mathcal{BR}(Z_c^\prime\to \etacrho)/\mathcal{BR}(Z_c^\prime\to \hcpi)$ (right). The red curve is the molecular prediction, whereas the black curve gives the predictions for dynamical type~I tetraquark model. The green (yellow) bands give the $68\%$ ($95\%$) confidence region. The pure kinematical tetraquark model predicts larger values, which separate even better from the molecular predictions.}
\label{fig:likelihood}
\end{figure}

Concerning the tetraquark pictures, we used both the type~I~\cite{maianiX} and type~II~\cite{maianitypeII} paradigms, with and without taking into account a recently developed dynamical picture~\cite{Brodsky:2014xia}. The outcome of this analysis is that all the considered models predict at least one of $Z_c$ and $Z_c^\prime$ to decay copiously into $\etacrho$. 
On the other hand, within the molecular model, we used NREFT~\cite{cleven} to evaluate the branching fractions. We found that the $Z_c^{(\prime)}\to\etacrho$ channels are strongly suppressed with respect to both $Z_c \to \jpsipi$ and $Z_c^\prime\to \hcpi$, and hence there is a fair probability for these decays not to be seen. 
In \figurename{~\ref{fig:likelihood}} we can appreciate indeed that the prediction for the $Z_c^\prime$ is well separated between tetraquark and molecular models. For the $Z_c$ the predictions are well separated in a type~I model, whereas molecular and type~II tetraquark model give solutions compatible at $2\sigma$ level. In the latter plot we include the dynamical estimates, but our results are even better separated if we consider the pure kinematical evaluation of the tetraquark decays.

Moreover, the molecular picture predicts $\mathcal{BR}(Z_c\to \hcpi)/\mathcal{BR}(Z_c^\prime \to \hcpi)<0.88$ and $\mathcal{BR}(Z_c\to\jpsipi)/\mathcal{BR}(Z_c^\prime\to\jpsipi)<1.86$ at 95\% C.L.. While the former seems in agreement with the experimental $\hcpi$ invariant mass distribution (where a hint of $Z_c$ is seen, albeit not statistically significative), it might be in contrast with the latter, which does not show any hint for $Z_c^\prime$. If the molecular hypothesis is correct, a higher statistics should make the $Z_c^\prime$ visible in $\jpsipi$. 

According to these results, the analysis of the $\etacrho$ final state could shed some light on the long-standing problem about the interpretation of the $Z_c$ and $Z_c^\prime$ states.

\section*{Acknowledgments}
We wish to thank A.A.~Alves~Jr., M.~Destefanis, R.~Faccini, A.D.~Polosa, F.~Piccinini and V.~Santoro for useful and interesting discussions.

\appendix
\section{Heavy meson chiral Lagrangians}
\label{appA}

We report here a brief review on heavy meson chiral Lagrangians in the case of the interaction with light vector mesons~\cite{Casalbuoni:1996pg}. The interaction Lagrangian is given in Eq.~\eqref{eq:Lrho}, where $H_{1a}$ and $H_{2a}$ are the Heavy Quark Effective Theory bi-spinors for mesons and anti-mesons respectively:
\begin{gather}
H_{1a}=\left(\frac{1+\slashed{v}}{2}\right)\left[V^\mu_a \gamma_\mu+P_a\gamma_5\right], \quad H_{2a}=\left[\bar V^\mu_a\gamma_\mu-\bar P_a\gamma_5\right]\left(\frac{1-\slashed{v}}{2}\right),\quad \bar H_{1,2a}=\gamma_0 H^\dagger_{1,2a}\gamma_0\\
\Psi=\left( \frac{1+\slashed{v}}{2}\right)  \left[\psi_\mu \gamma^\mu - \eta_c \gamma^5\right]  \left( \frac{1-\slashed{v}}{2}\right), \quad \chi_\mu=\left( \frac{1+\slashed{v}}{2}\right)  \left[\cdots + h_{c,\mu} \gamma^5\right]  \left( \frac{1-\slashed{v}}{2}\right), \quad \mathcal{Z}^{(\prime)}_\mu=\left( \frac{1+\slashed{v}}{2}\right)  Z^{(\prime)}_\mu \gamma^5  \left( \frac{1-\slashed{v}}{2}\right),
\end{gather}
where $v_\mu$ is the heavy meson velocity and $a$ is a flavor index. The fields $V^\mu$ $(\bar V^\mu)$ and $P$ $(\bar P)$ annihilate a (anti-)vector and a (anti-)pseudoscalar respectively according to $V^\mu\left|V(q,\epsilon)\right\rangle\!=\!\epsilon^\mu\sqrt{M_V}\left|0\right\rangle$, 
$P\left|P(q)\right\rangle\!=\!\sqrt{M_P}\left|0\right\rangle$. In the definition of $\chi_\mu$, we omit the $C=+1$ $P$-wave charmonia. 
For light vector mesons, we follow the anti-hermitian convention of~\cite{Casalbuoni:1996pg}, {\it i.e.} $\rho_\mu=i g_V\hat \rho_\mu/\!\sqrt{2}$, where:
\begin{equation}
 \hat \rho_\mu =\left( \begin{matrix}
                  \rho^0_\mu/\!\sqrt{2} + \omega_\mu/\!\sqrt{2}  & \rho^+_\mu\\
                  \rho^-_\mu & -\rho^0_\mu/\!\sqrt{2} + \omega_\mu/\!\sqrt{2}
                 \end{matrix}\right),
\end{equation}
and $g_V\simeq 5.8$~\cite{Casalbuoni:1996pg,Isola:2003fh}.
Starting from the Lagrangians \eqref{eq:Lrho} one can obtain the Feynman rules for different processes, which we report in \tablename{~\ref{tab:rules}}. The non-relativistic limit can be obtained by letting $v\to (1,\vec 0)$.

\section{Amplitudes and loop integral}
\label{appB}According to the Feynman rules found in \tablename{~\ref{tab:rules}} and in~\cite{cleven} the non-relativistic one loop amplitude associated with the processes in \figurename{~\ref{fig:one_loop}} is:
\begin{subequations}
\begin{align}
\mathcal{A}_{Z_c\eta_c\rho}&=2\sqrt{2M_ZM_\eta}\,zg_Vg_2\,\Bigg[\beta \frac{\vec q_\rho\cdot \vec \eta}{m_\rho}\lambda^i\left(I^i(M_D,M_{D^*},M_D;q_\rho)+I^i(M_{D^*},M_D,M_{D^*};q_\rho)\right) \\
&\quad+ \lambda \left((\vec\lambda\cdot\vec \eta)q_\rho^i-(\vec\lambda\cdot\vec q_\rho)\eta^i\right)\left(I^i(M_D,M_{D^*},M_{D^*};q_\rho)+I^i(M_{D^*},M_D,M_{D^*};q_\rho)\right)\Bigg],
\end{align}
\end{subequations}
where $\vec \lambda$ and $\vec \eta$ are the spatial polarizations of the $Z_c$ and of the $\rho$ respectively and $q_\rho$ is momentum of the outgoing $\rho$. The overall factor of 2 comes from the charge conjugate diagrams. $I^i(m_1,m_2,m_3;q)$ is the non-relativistic loop integral with three propagators:
\begin{subequations}
\begin{align}
I^i(m_1,m_2,m_3;q)=&\frac{i}{8}\int \frac{d^4l}{{(2\pi)}^4}\frac{{(q-2l)}^i}{\left(l^0-\frac{\vec l^2}{2m_1}-m_1+i\epsilon\right)\left(M_{Z^{(\prime)}}-l^0-\frac{\vec l^2}{2m_2}-m_2+i\epsilon\right)\left(l^0-q^0-\frac{{(\vec l-\vec q)}^2}{2m_3}-m_3+i\epsilon\right)}.
\end{align}
It can be computed with the same techniques explained in \cite{cleven} and it gives:
\begin{align}
I^i(m_1,m_2,m_3;q)=q^i\left[I_0(m_1,m_2,m_3;q)-2I_1(m_1,m_2,m_3;q)\right],
\end{align}
with:
\begin{subequations}
\begin{align}
I_0(m_1,m_2,m_3;q)&=\frac{\mu_{12}\mu_{23}}{16\pi\sqrt{a}}\left[\tan^{-1}\left(\frac{c_{23}-c_{12}}{2\sqrt{ac_{12}-i\epsilon}}\right)+\tan^{-1}\left(\frac{2a+c_{12}-c_{23}}{2\sqrt{a(c_{23}-a)-i\epsilon}}\right)\right], \\
I_1(m_1,m_2,m_3;q)&=\frac{1}{2a}\left\{\frac{\mu_{12}\mu_{23}}{2}\left[B(c_{23}-a)-B(c_{12})\right]+(c_{23}-c_{12})I_0(m_1,m_2,m_3;q)\right\}.
\end{align}
\end{subequations}
In particular, $\mu_{ij}$ is the reduced mass between $m_i$ and $m_j$, $a=\left(\frac{\mu_{23}}{m_3}\right)^2q^2$, $c_{12}=2\mu_{12}(m_1+m_2-M_{Z^{(\prime)}})$ and $c_{23}=2\mu_{23}\left(m_2+m_3+q_0+\frac{\vec{q}^2}{2m_3}-M_{Z^{(\prime)}}\right)$. Lastly:
\begin{align}
B(c)=-\frac{\sqrt{c-i\epsilon}}{4\pi}.
\end{align}

\end{subequations}
For the case of the $Z_c^\prime$ we have, instead:
\begin{subequations}
\begin{align}
\mathcal{A}_{Z_c^\prime\eta_c\rho}&=2\sqrt{2M_{Z}M_\eta}z^\prime g_2g_V\,\Bigg[\lambda\left( (\vec\lambda\cdot\vec\eta)q_\rho^i -(\vec\lambda\cdot\vec q_\rho)\eta^i\right)\left(I^i(M_{D^*},M_{D^*},M_D;q_\rho)+I^i(M_{D^*},M_{D^*},M_{D^*};q_\rho)\right) \\
&\quad+2\beta\frac{\vec q_\rho\cdot\vec \eta}{m_\rho}\lambda^iI^i(M_{D^*},M_{D^*},M_{D^*};q_\rho)\Bigg].
\end{align}
\end{subequations}





\begin{table}[t]
\centering
\begin{tabular}{c|c|c}
\hline\hline 
Process & Relativistic rule & Non-relativistic rule \\
\hline\hline
$D(P)\to\rho(k,\eta)D(q)$ & $-i\sqrt{2}M_D\beta g_V(v\cdot \eta^*)$ & $-i\sqrt{2}M_D\beta g_V(\vec k\cdot \vec\eta^*/m_\rho)$ \\
\hline
$D(P)\to\rho(k,\eta) D^*(q,\epsilon)$ & $-\sqrt{2M_DM_{D^*}}g_V\lambda \epsilon^{\mu\nu\rho\sigma}v_\mu \epsilon^*_\nu k_\rho \eta^*_\sigma$ & $\sqrt{2M_DM_{D^*}}g_V\lambda \epsilon^{ijk} k^i\eta^{*j} \epsilon^{*k}$ \\
\hline
\multirow{3}{*}{$D^*(P,\lambda)\to \rho(k,\eta) D^*(q,\epsilon)$} & $i\sqrt{2}M_{D^*}g_V\beta(\epsilon^*\cdot\lambda)(v\cdot\eta^*)+$ & $-i\sqrt{2}M_{D^*}g_V\beta(\vec \epsilon^*\cdot\vec \lambda)(\vec k\cdot\vec \eta^*/m_\rho)+$ \\
& $+i\sqrt{2}M_{D^*}g_V\lambda\left[(\epsilon^*\cdot\eta^*)(\lambda\cdot k)-\right.$ & $+i\sqrt{2}M_{D^*}g_V\lambda\left[(\vec \epsilon^*\cdot \vec k)(\vec \lambda\cdot \vec \eta^*)-\right.$ \\
& $\left. -(\epsilon^*\cdot k)(\lambda\cdot \eta^*)\right]$ & $\left.-(\vec \epsilon^*\cdot\vec \eta^*)(\vec \lambda\cdot \vec k)\right]$ \\
\hline
\end{tabular}
\caption{Feynman rules involving the $\rho$ meson. The charge conjugated of all rules gain an additional minus sign.} \label{tab:rules}
\end{table}

\end{document}